# Field-induced reversible phase transition and negative differential resistance in In$_2$Se$_3$ ferroelectric semiconducting FETs

Jishnu Ghosh[1,#], Shubham Parate[1,#], Arup Basak[1], Binoy Krishna De[1], Krishnendu Mukhopadhyay[1], Abhinav Agarwal[1], Gopesh Kumar Gupta[2], Digbijoy Nath[1]* & Pavan Nukala[1,2]*

[1] Center for Nanoscience and Engineering, Indian Institute of Science, Bengaluru, India

[2] Department of Materials Engineering, Indian Institute of Science, Bengaluru, India

* Corresponding authors' email

pnukala@iisc.ac.in, digbijoy@iisc.ac.in

[#] These authors contributed equally

## Abstract

Indium selenide (In$_2$Se$_3$), a ferroelectric semiconductor, offers a unique platform for multifunctional nanoelectronics owing to the interplay between polarization dynamics, interlayer sliding, and structural polymorphism. Ferroelectric semiconductor field-effect transistors (FeS-FETs) provide an ideal architecture to harness this coupling. Here, we demonstrate gate-tunable negative differential resistance (NDR) with high peak-to-valley ratios and hysteretic output conductance in In$_2$Se$_3$ FeS-FETs. Combining high-resolution electron microscopy with electrical transport measurements, we attribute the NDR to a field-induced, volatile phase transition from a low-resistance α-2H phase to a high-resistance state. Atomic-scale ex-situ imaging reveals that in-plane electric fields (V$_d$) drive interlayer sliding, rotational misalignments that generate Moiré patterns, and intralayer shear—together producing stress-induced phase transitions. Out-of-plane field however results in robust non-volatile polarization switching. These mechanistic insights highlight both the promise of two-dimensional ferroelectric devices for multifunctional nanoelectronics and alternative computing paradigms, and the intrinsic limitations of In$_2$Se$_3$ field-effect transistors for conventional ferroelectric memory applications.

**Introduction**

In the pursuit of materials that can anchor the next generation of intelligent, adaptive electronics and optoelectronics, In$_2$Se$_3$ has emerged as a uniquely versatile contender. It is one of the rare systems that intrinsically integrates semiconducting functionality with robust ferroelectricity at an atomic scale exhibiting both out of plane (OOP) and possible in-plane (IP) spontaneous polarization[1–6]. With a moderate bandgap of approximately 1.3 to 1.4 eV,[7] along with an air stable van der Waals layered crystal structure, stackable into various polymorphs, In$_2$Se$_3$ offers unique combination of properties which can be exploited to design new functional devices.

For example, ferroelectric semiconducting field effect transistors (FeS-FETs) which offer larger memory window at lower supply voltages compared to conventional FeFETs were realized on In$_2$Se$_3$ platform[8–11]. Gate triggered polarization induced resistive switching in FeS-FETs was used to implement synaptic device functionalities, especially in context of pattern recognition[12–14] and memristive behavior[15]. In-plane voltage gated co-planar FETs with good on-off ratios were also shown on heterostructures of In$_2$Se$_3$ and MoS$_2$ layers, which demonstrated finer electric field control of ferroelectricity.[16] By employing other innovative electrical device geometries, novel devices such as switchable diodes [17,18], and functionalities such as rectification [19,20], and high-density memory [21,22] have been demonstrated on this material platform. The optoelectronic properties of In$_2$Se$_3$ were harnessed in devices including high-performance visible and infrared photodetectors[10,23,24], as well as optical synapses for artificial vision systems.[25]

In$_2$Se$_3$ exhibits rich polymorphism with several metastable phases which are energetically close to each other[26–28]. All the phases are characterized by quintuple layer of Se-In-Se-In-Se bonded to other quintuples through van der Waals bonding. The intralayer and the interlayer stacking sequence together determine the exact phase and symmetry of the material. The commonly reported phases are α-In$_2$Se$_3$ which comprise of polar 2H or 3R phases[16], non-polar β phases[17,18,29,30], polar β' phases comprising of 1T stacking[26]. Earlier works largely attributed novel device behaviours to polarization switching in conjunction with semiconducting nature of In$_2$Se$_3$[2,31]. However, recent findings increasingly highlight the role of external stimuli in triggering structural modifications and their dynamic effects on the electronic and optical behavior of the devices[26,31–35].

For instance, in-situ electron microscopy studies showed that in-plane electric field creates sliding faults and coupled domain boundaries in In$_2$Se$_3$ nanowire devices. Interaction between the domain boundaries leads to electromechanical shocks (Barkhausen's noise[36]), which leads to amorphization of In$_2$Se$_3$ in a completely solid-state manner. Such a crystal-amorphous transition was exploited to create low-power phase change memory[26,37]. Reversible phase change memory functionality was also exploited in optoelectronic context to create electrochromic devices[38].

In another work, a current (and device temperature) controlled unzipping-zipping 2H-α to 2H-β phase transition and pathways were visualized through in-situ electron microscopy[39]. In-situ STEM studies also revealed an electrically induced domain wall motion in various phases, and a large, field-induced ferro to paraelectric transitions in 2H and 3R phases through intralayer atomic gliding[40]. Femtosecond laser induced phase transitions were also recently reported in In$_2$Se$_3$[34]. Insights into these dynamic phase transitions driven by electric field, light, stress, and temperature enable the design of more versatile adaptable and multifunctional devices [12,13,25,32–34].

Here, we show that in-plane field driven structural dynamics such as layer sliding, layer rotation and Moiré pattern formation, and associated stress-induced phase transitions enable a robust gate tunable hysteretic negative differential resistance (NDR) response with good peak to valley ratios in FeS-FET devices. The out-of-plane field, however, results in robust ferroelectric switching of these devices. These novel functions open up $In_2Se_3$ in the context of multivalued logic and synaptic devices, neuromorphic oscillators and associated computing paradigms, albeit with a better structural understanding.

**Results**

Fig 1a illustrates the schematic architecture of the back-gated ferroelectric semiconductor field-effect transistor (FeS-FET), while the corresponding optical microscopy (OM) image of a representative device is shown in Fig 1b. The device was fabricated using mechanically exfoliated flakes of α (2H) phase $In_2Se_3$ transferred onto a $Si/SiO_2$ (90 nm) substrate. Source and drain electrodes consisting of Cr/Au (10/30 nm) were patterned by standard e-beam lithography, with a channel length of ~3 μm. The use of a global back gate allows electrostatic control across the entire $In_2Se_3$ channel. The flake thickness was confirmed to be ~35-40 nm by atomic force microscopy (Fig S1a and b).

High angle annular dark field scanning transmission electron microscopy images (HAADF-STEM) of the virgin state channel region obtained in both cross-section (zone axis: a axis) and plan view (zone axis: c axis) are presented in Fig 1c and 1d. Both the views show unit cells (identified in the respective insets), consistent with 2H phase of $In_2Se_3$ (also see Raman spectra in Fig S2). Unit cell in the cross-sectional view extends over two successive quintuple layers of Se-In-Se-In-Se stacked as *abcca baccb*. Such an asymmetric stacking is responsible for out of plane polarization, which is mapped in Fig S3a, with an average out of plane displacement of Se sublattice with respect to In sublattice of ~50 pm (see Supplementary note 1, Fig S3a and c for histogram of displacements). Also noteworthy is the presence of a non-zero net in-plane displacement (~ 6-7 pm), along the $<\bar{1}2\bar{1}0>$ axis (b' axis) (Fig S3a). Plan view images also reveal a similar ~ 6-7 pm displacement along both a and b' axes (see Fig S3b, d).

To measure the OOP ferroelectric switching, we performed both piezoforce microscopy (PFM) as well as Polarization-Electric field (P-E) measurements on ferroelectric capacitors., PFM was conducted in dual AC resonance tracking mode (DART-PFM) on flakes exfoliated onto a gold-coated Si wafer, which served as the back electrode (Fig 1e). A train of voltage pulses from –20 V to +20 V was applied through a conductive Ti tip in contact with the flake surface, and the piezoelectric response was measured in between these pulses in an OFF-state. The measured amplitude response exhibited the characteristic butterfly-shaped curve (Fig 1e), while the phase signal revealed a 180° reversal (Fig 1e), out-of-plane (OOP) ferroelectric polarization switching.

FE capacitors were fabricated with $In_2Se_3$ flakes embedded in hBN (to passivate the interfaces) electroded by multilayer graphene as both top and the bottom contacts (Graphene/hBN/$In_2Se_3$/hBN/graphene, schematic shown in Fig S4). P-E loop (Fig 1f) of a representative device shows an effective remnant polarization ($P_r$) of 2.7 μC/cm², and a very low coercive field ($E_c$) of 0.3 MV/cm. Note that the presence of hBN capping layer renders an effective device $P_r$ which is less than $P_r$ of $In_2Se_3$ layer, and an effective $E_c$ which is larger. Nevertheless, it is important to note that this is the first time $P_r$, $E_c$ and ferroelectric switching are reported through direct measurements on (macroscopic) capacitor devices in semiconducting $In_2Se_3$.

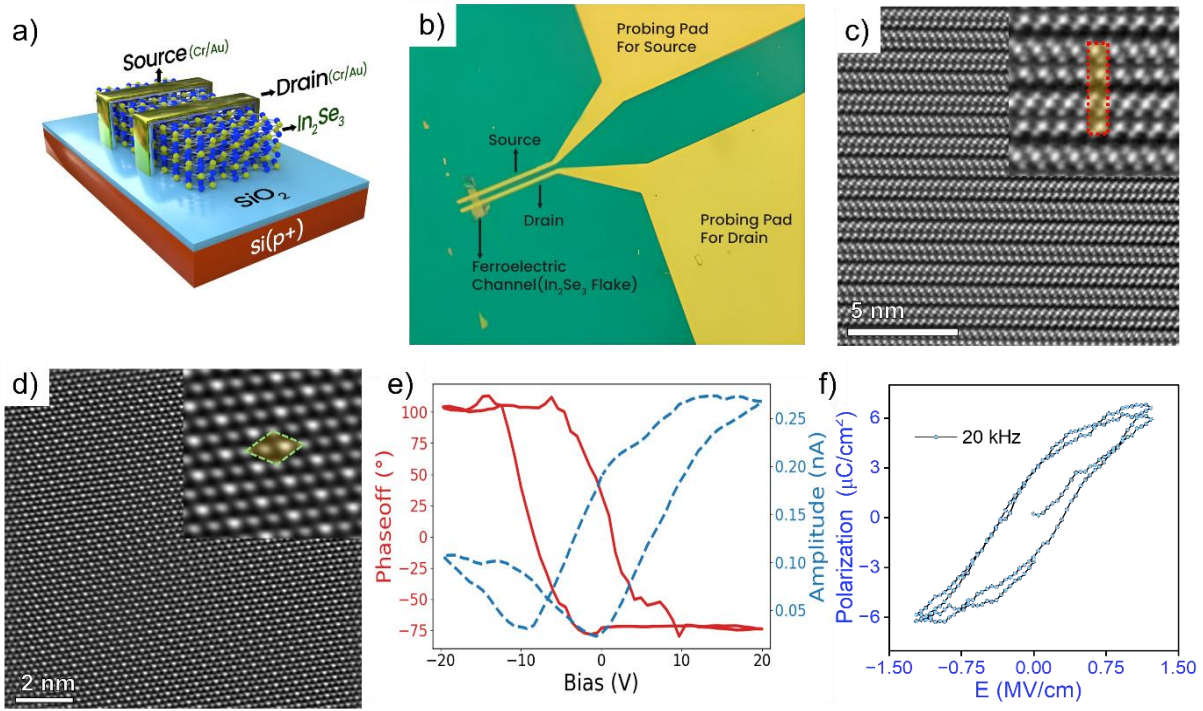

**Fig 1. Device structure and characterization:** a) Schematic of a back-gated ferroelectric FET (FeS-FET) with α-In₂Se₃ as the channel. b) Optical image of a fabricated FeS-FET on heavily doped Si with 90 nm SiO₂; with ~3 μm channel. HAADF-STEM image showing c) cross section of In₂Se₃, and d) plan view, both consistent with In₂Se₃ in a layered 2H phase. e) PFM amplitude vs. bias showing butterfly hysteresis loop and PFM phase response with 180° shift, confirming strong out-of-plane ferroelectricity in In₂Se₃. f) PE loop measured on multilayered graphene/hBN/In₂Se₃/hBN/multilayered graphene capacitors in a top-down geometry (see Fig S4 and supplementary note 3).

Fig 2a shows the output characteristics ($I_d - V_d$) of the device measured at varying gate voltages where the drain voltage was swept from 0 V to 10 V. A pronounced negative differential resistance (NDR) behaviour is observed at all gate voltages (Fig 2a), with the peak-to-valley current ratio (PVCR) becoming more distinct as $V_g$ increases (Fig 2a), and this is consistent across various devices (Fig 2c). Such PVCRs were reported on FETs of other non-ferroelectric 2D platforms only at fields that are atleast three times as large[41]. NDR and PVCR diminish with increasing negative gate voltage (Fig 2b). Furthermore, onset of NDR ($dI/dV = 0$, maxima in $I_d - V_d$) also shifts to higher $V_d$ with increasing $V_g$ (Fig S5a and b), and this aspect will be elaborated upon later (also see supplementary note 4). It is interesting to note that our virgin devices exhibited NDR behavior at $V_g = 0$ V in Fig 2b, suggesting that a significant effect arises from drain voltage (in-plane field) induced changes to the material.

Next, we checked if the NDR behavior in $I_d - V_d$ characteristics in any particular device is repeatable. Fig 2d shows data on a device subjected to five consecutive cycles of drain voltage ($V_d$) swept from 0 to 10 V and back, at $V_g = 15$ V. A clear clockwise hysteretic behaviour is observed in every cycle, with NDR only present in the forward sweep, and the $I_d$ in the reverse sweep lower than that in the forward sweep. Our data suggests (with more structural proof ahead) that the NDR arises from a $V_d$ induced volatile phase transition of $In_2Se_3$ from a low resistance state (LRS) to a higher resistance state (HRS). In subsequent cycles, the device resets back to LRS, and the NDR and hysteretic behaviour repeats. This suggests that HRS relaxes back to LRS upon removing the field within certain timescale.

In order to understand the time dependent hysteretic nature of our devices, we performed $I_d - V_d$ sweeps (at $V_g = 5$ V) by varying the $V_d$ step size from 50 mV to 500 mV (Fig 2e). At larger step sizes (faster measurements), we see that the $I_d$ is higher (and NDR is prominent), suggesting slower charge trapping and de-trapping events at play. The timescales of these processes were determined through constant DC voltage measurements performed on devices before onset of NDR ($V_d = 3$ V), and after the onset NDR ($V_d = 4.5$ and 9 V) (Fig S6a-c, also see supplementary note 5). We estimate relaxation time constants ($<\tau_r>$) ~8 sec subsequent to NDR, which reduces to 2-2.5 sec before onset of NDR. The change in these time constants lend support to the hypothesis that NDR represents a field induced reversible phase transition from LRS to HRS, both phases distinguishable through their relaxation time constants. Given that

HRS possesses a long $\tau_r$ (8 sec), this transition appears non-volatile and hysteretic when forward and reverse sweep are performed fast enough (compared to time scale set by $\tau_r$) and show a clear dependence on the time steps used in the quasistatic $I_d - V_d$ sweeps. In other words the LRS to HRS transition has a short-term memory of $\tau_r$ (with non-volatility within $\tau_r$) and is a volatile transition beyond $\tau_r$. In addition, the slight cycle to cycle fluctuations (for example, peak NDR current reduces from ~30 to 25 µA from cycle 1 to cycle 4, (Fig 2d)) reflects some irreversibility and remanence of this phase transition.

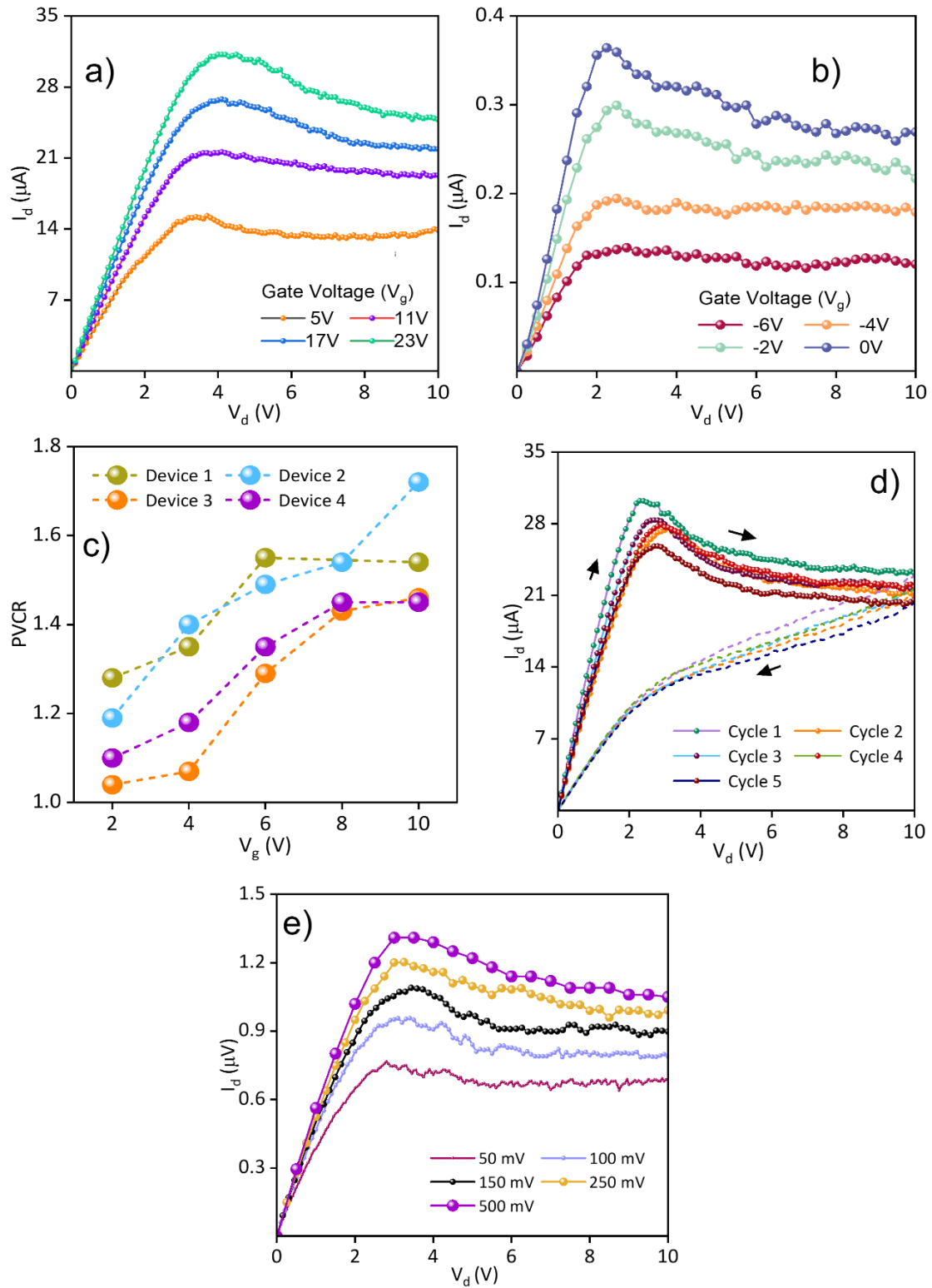

**Fig 2. Electrical performance and NDR behavior of In₂Se₃ FeS-FET:** a) FeS-FET Output characteristics ($I_d – V_d$) with increasing gate voltages ($V_g$ = 5 to 23 V). b) Output characteristics under negative gate bias ($V_g$ = −6 to 0 V). c) PVCR vs. gate voltage for four devices (D.1–

D.4). d) Cycling effect on $I_d - V_d$ sweep under $V_g = 15$ V. e) Step-size (of $V_d$) dependent $I_d - V_d$ characteristics at fixed gate voltage ($V_g = 15$ V).

The LRS to HRS transition induced by voltage, can either be a result of either field-induced or temperature (Joule heating) induced phase transitions. Ferroelectric to paraelectric reversible phase transitions were reported in $In_2Se_3$ beyond ~260 °C [42] of device temperature. It is possible that our devices heat up beyond the phase transition temperatures to affect a volatile LRS to HRS transition, and the relaxation timescales in seconds can be a result of long thermal time constants of the system.

In order to understand the effect of device temperature, we first performed temperature dependent $I_d - V_d$ measurements at $V_g = 0$ V (Fig 3a), from 77 K to room temperature. NDR is observed at all the temperatures, and the VI power input into the device at the onset of NDR shows a decreasing trend (Fig S7, see supplementary note 6) with increase in ambient temperature. We then estimated the maximum device temperature under operating conditions of various devices through electrothermal simulations using COMSOL, presented in Supplementary note 7 (Fig S8 and S9). Overall, device steady-state temperatures compared to ambient (ΔT) increase by < 0.02 °C (Fig 3b), ruling out the effect of any thermal phase transitions in our devices.

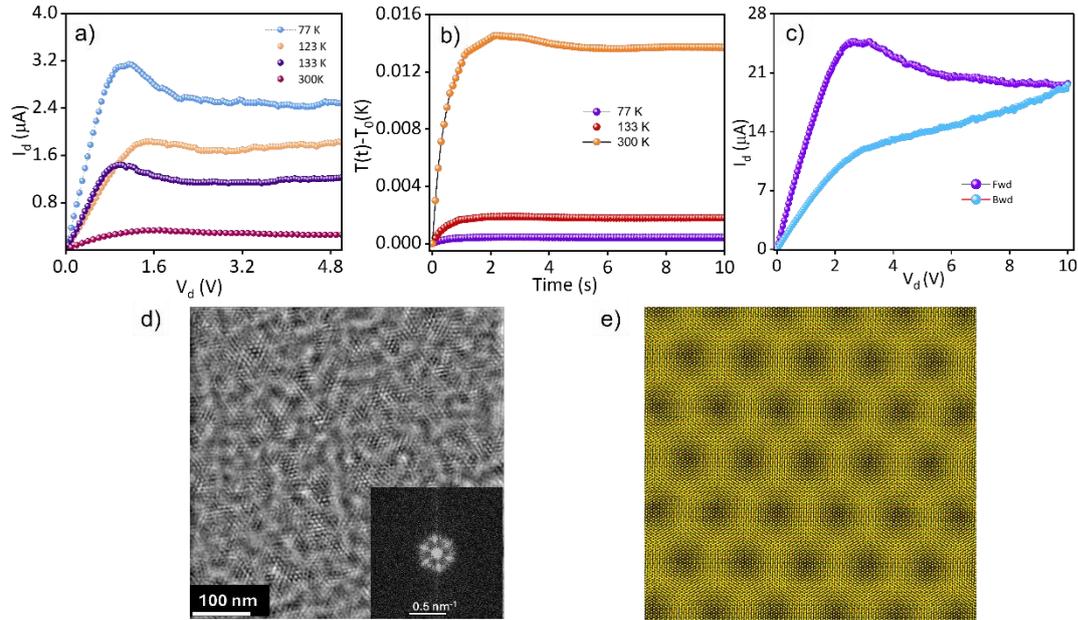

**Fig 3. Effect of temperature on device and structural analysis of channel:** a) Temperature dependent $I_d$ – $V_d$ measurements at zero gate bias. b) COMSOL simulated device steady state temperature change with time. c) Applied I-V characteristics at $V_g = 0$ V for the device on which TEM was subsequently performed. d) Large area Moiré pattern and its Fast Fourier Transform (FFT, inset). e) Schematic showing Moiré pattern by rotating two layers by an average angle of 2.66° calculated from Moiré periodicities observed in (d).

Recall that our devices show a cycle-to-cycle reduction in overall drain current (Fig 2d). The remnant structural feature responsible for these permanent changes, also give us indications into the nature of volatile (beyond $\tau_r = 8$ sec) field-induced phase transitions captured in transport measurements. We analysed the bias-induced structural changes through HAADF-STEM imaging. The virgin state structure is presented in Fig 1c and d.

We then cycled the same device two times through NDR at $V_g = 0$ V (Fig 3c). In the plan-view (Fig 3d, higher resolution image is shown in Fig 10), we observe formation of Moiré fringes, clearly suggesting irreversible layer rotation (and sliding) due to the applied field. From the Moiré periodicity (obtained from FFT (Fig 3d, inset), and the in-plane lattice parameters of $In_2Se_3$, we infer a field induced average interlayer twist angles of upto 2.66°, with a schematic of Moiré pattern shown in Fig 3e (see Supplementary note 8). The diffuse nature of the FFT (Fig 3d, inset) suggests a significant spread of the twist angle about the average value.

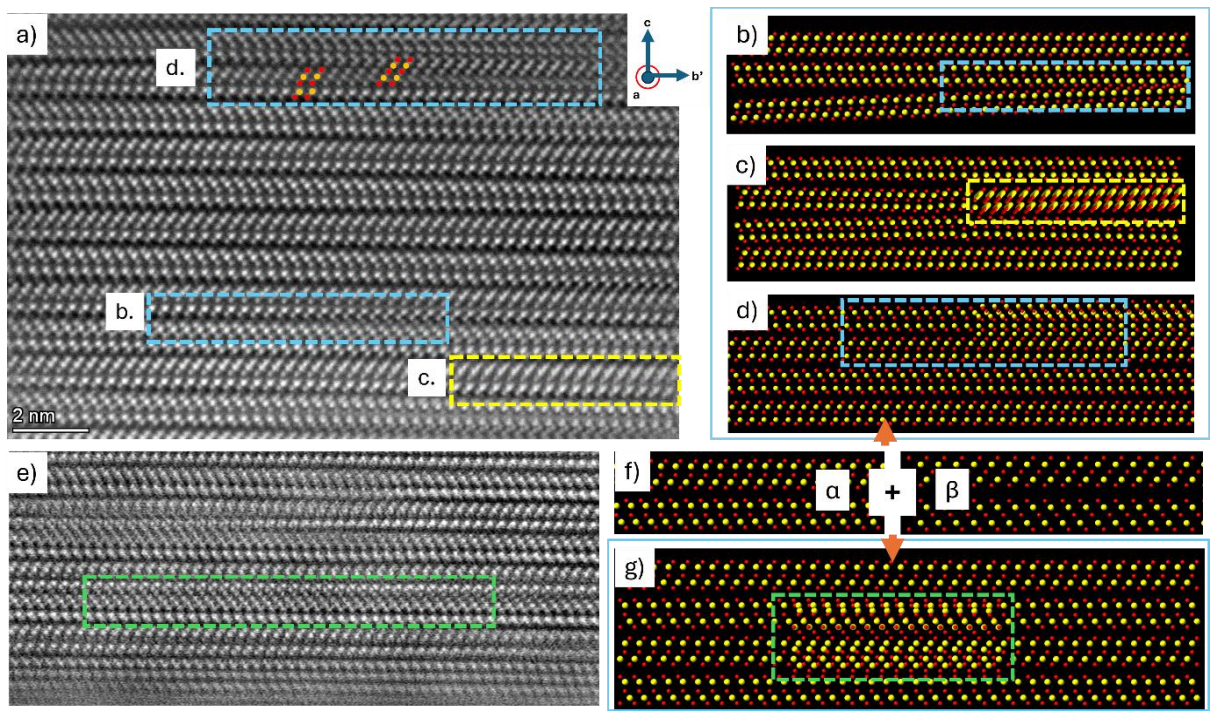

**Fig 4. Field-induced structural changes:** (a) Shows the HAADF image along <10$\bar{1}$0> zone with some defective regions marked as b, c, d. Overlay of atoms shows stacks of α (left) and β-like (right) phases, suggesting that region d shows an overlap of α and sheared β-like phases. *a, b, c* directions are shown in inset. (b) Schematic created through out of plane tilt (or bend) of the bottom quintuple layer (about the *a*-axis), with respect to rest of the crystal replicating zipped-unzipped regions marked in Fig 4a as b. (c) shows a schematic of an image created by a combination of in-plane twist, and out-of-plane shear (and tilt) of a quintuple layer shown in yellow box, replicating a region marked as c in Fig 4a. (d) Schematic explaining formation of region d in Fig 4a, as a combination of α and sheared β/ β'-like phases. e) Another HAADF image from a different region showing similar defective features. f) Schematic of α-phase and β-phases. (g) A schematic overlay of α-phase and sheared β/ β'-like phases, representing green boxed region in Fig 4e.

Layer sliding and rotations can further be gleaned from cross-sectional images. A representative cross-sectional image is shown in Fig 4a, which shows several defective regions. We show that each of these regions can be formed as one or a combination of in-plane rotations, out-of-plane tilting, out-of-plane shear and frozen incipient phase transition fronts, all of which occurring due to piezoelectric stress caused due to in-plane electric field. For e.g. schematic in Fig 4b is created through out of plane tilt (or bend) of the bottom quintuple layer (about the *a*-axis), with respect to rest of the crystal. This operation leads to zipped-unzipped regions marked in Fig 4a as b. Fig 4c shows a schematic of an image created by a combination of in-plane twist, and out-of-plane shear (and tilt) of a quintuple layer shown in yellow box. This is a good replication of a region marked as c in the HAADF-STEM image shown in Fig 4a. Fig 4d represents resultant overlap between an α phase (with quintuple layer stacking sequence: *abcca*) and β (β') phase (quintuple layer stacking sequence: *abcab*), with the β (β') phase slightly displaced (out-of-plane shear) from the α phase along the c-axis (b'-c plane). This replicates region (d) in Fig 4a, where the quintuple layer stacking of the α and β (β') phases are marked (also shown in Fig 4f). Fig 4e shows another HAADF-STEM image, with a region of interest marked in green box, which again can be interpreted as of α and β (β') phases overlaid with some out-of-plane shear (Fig 4g). More images in other cycled devices with in-plane twist and incipient phase transitions are shown in Fig S10 and S11.

We identify angles θ1- θ4 in a particular quintuple layer as quantifiable indicators for various phases in $In_2Se_3$ (Fig S11 c,d). Table S3 shows the values of these indicators for quintuple layer orientations in several polymorphs of $In_2Se_3$. A particular angle of interest is θ2, which in the α (2H) phase is 90º and sharply reduces to ~60º in β/β' phases (see supplementary note 9 – 10). In cycled device, while most regions remain in α (2H) phase as determined from these indicators (R3-R6 in Fig S11b) regions marked as R7 and R8 in Fig. S11b and Fig. S10 respectively are consistent with the β' phase (Fig S11f). As a result, the NDR most likely represents a field induced reversible transition between α and β' phases.

**Discussion**

α to β transition, in particular, has been shown on 2 terminal $In_2Se_3$ flakes via in-situ microscopy experiments[39], and is indeed preceded by zipping-unzipping structural modifications. However, due to the lack of effective heat sinking in the in-situ TEM device geometry reported[39], the applied current caused the device temperature to rise above the ferroelectric–paraelectric transition temperature (260 - 270 °C)[42], thereby inducing the phase transition. Furthermore, electrically these devices did not show any NDR behavior. Furthermore, there are some reports which suggest that β phase is more conducting than the α phase [39,42,43], which means that α to β phase transition (with increasing voltage) cannot give N-shaped NDR.

In this work, however, we show through electrothermal simulations that increase in device temperature is negligible (because of efficient heat sinks), and the observed phase transition is solely a field-dependent transition. NDR signifies a field-induced transition between LRS α-2H phase to possibly a more resistive (HRS) β' phase (to be addressed as β'-like phase from here on). This transformation is preceded by layer sliding, rotations and out-of-plane shear and bending, all of which is triggered by the piezoelectric stress in the system. Note that while field-induced α to β'-like transition presented in this work is mostly reversible, certain irreversible features (and corresponding defects) of transition such as slid layers (sliding faults), rotated layers (Moiré patterns), sheared planes (stacking faults) and some locally quenched (β'-like/α) phase transition fronts still remain and possibly affect the cyclability of FeS-FETs.

α to β'-like phase transition is a polar to non-polar phase transition and can be tuned with electric field. In typical ferroelectrics, electric field enhances the stability of the polar phase, increasing the activation barrier for the formation of the non-polar phases. In case of $In_2Se_3$, positive gate voltage provides OOP electric field, that increases the "up" polarization towards saturation in an already "up" polarized virgin state (Fig S3a). This stabilizes the polar α-phase and manifests as increase of drain voltage for the onset of NDR (or LRS to HRS) in the accumulation region (Fig S5b, Fig 2a). The unique aspect in $In_2Se_3$ (and we believe in 2D ferroelectrics in general) is that it is the only the out-of-plane field that stabilizes the polar phase, but in-plane field favors transitions to non-polar phase(s). In-plane field (through piezoelectric stress and plastic deformation) has the effect of disordering both structure and polarization, similar to the effect of temperature. The complete effect of gate voltage on the channel conductivity is shown in the transfer characteristics in Fig S12e and f, and discussed further in supplementary note 11.

Finally, our TEM results also shed light on the structural features responsible for long time constants in HRS. It is by now clear that the HRS phase contains all the field irreversible defects (Moiré boundaries, sliding faults, shear faults, quenched phase fronts) in addition to field reversible defects, all of which can act as charge trapping sites giving rise to very slow charge dynamics (~8 sec). An LRS phase, in contrast contains only field irreversible defects, and this explains the reduction in the time constant to ~2 sec. All in all, the robust out-of-plane non-volatile ferroelectric transition, together with the volatile (yet relatively slow) in-plane dynamics, makes our devices highly attractive for exhibiting neuromorphic behavior—encompassing both short-term (on the order of seconds) and long-term memory—within a single device platform.

**Conclusions:**

We report a field-induced, gate-tunable, volatile phase transition (low-resistance state, LRS, to high-resistance state, HRS) driven by an in-plane electric field in α-$In_2Se_3$ FeS-FET devices, which manifests as a negative differential resistance behavior. The phase transition is triggered by piezoelectric stress and is preceded by in-plane layer-sliding rotations and out-of-plane shearing of the quintuple layers. Despite their volatility, the HRS and LRS exhibit carrier relaxation time constants of ~8 s and 2–2.5 s, respectively, making them attractive for implementing neuromorphic functionalities. Our findings reveal that while out-of-plane switching in In$_2$Se$_3$ remains robust, the in-plane field-driven dynamics are more complex than simple polarization reversal. These structural dynamics render In$_2$Se$_3$ FeS-FETs less suitable for conventional memory applications in FET geometries, but highly promising for alternative computing paradigms that exploit controlled responses under multiple stimuli.

## Methods

**Device Fabrication**: α-In$_2$Se$_3$ flake was transferred onto 90 nm SiO$_2$ on a Si substrate using Scotch tape-based exfoliation and PDMS based dry transfer method. The p+ Si wafers with 90 nm thermally grown SiO$_2$ were prepared by dry oxidation method from commercially purchased Si wafers. After that source and drain contacts were patterned on the transferred flake by standard lithographic process. Then 10 nm Cr and 30 nm Au were deposited by electron-beam evaporation and followed by a liftoff process for α-In$_2$Se$_3$ back-gate transistor. PFM samples were prepared by exfoliating In$_2$Se$_3$ flakes and transferring it to gold deposited Si/SiO$_2$ substrate. Heterostructure for PE loop measurement was made by stacking vDW layers through PDMS based dry transfer method. Later e-beam lithography and subsequent metal deposition (Cr/Au-20nm/80nm) was done to define the electrodes.

**Device characterization**: The thickness of the α-In$_2$Se$_3$ was measured using a Park Systems NX20 AFM system using a non-contact cantilever. TEM analyses of the channel were done using a TEM- Titan 300 Themis D 3391. PFM measurements were performed on In$_2$Se$_3$ flakes transferred on Gold-coated Si/SiO$_2$ wafer using a Pt/Ti tip under a contact and dual-AC resonance mode. DC electrical characterization was performed with a Keysight Semiconductor Device Analyzer B1500A in a dark environment. Electrical data were collected with a Cascade

PM5 probe station at room temperature The device was cooled to 77 K using a liquid nitrogen cryostat (Lake Shore Cryotronics) for low-temperature electrical characterization. Ferroelectric P-E loops were measured using Radiant Multiferroic System.

**TEM sample preparation, experimental details:** Cross section as well as plan-view $In_2Se_3$ flakes were imaged in HAADF STEM mode by using low probe currents ~20 pA at 300 kV acceleration voltage using Thermofisher TITAN THEMIS 300 TEM equipped with $C_s$ aberration corrector. The convergence angle was 24.5 mrad with a HAADF collection angle of 48-196 mrad.

All biasing experiments were done ex-situ *i.e.* biasing was performed in DC probe station and later a device cross section lamella was prepared (or in case of plan view directly imaged) and then imaged in TEM. $In_2Se_3$ flakes were mechanically exfoliated from single crystal from HQ Graphene using scotch tape onto a PDMS sheet and then it was transferred on the plasma cleaned in-situ DENS MEMS biasing chip at 105 ºC. The plasma cleaning was done under 0.25 mbar oxygen pressure for 2 mins just before the transfer. Later FIB platinum was deposited to ensure proper contact.

Cross section lamella of flakes was made by FIB method from transistor devices (fabricated on $In_2Se_3$ flakes transferred on $SiO_x$ substrates) that showed NDR behavior while pristine cross section was taken from single crystal. A protective platinum deposition by electron beam and subsequently by ion beam platinum was done across source and drain followed by trenching, lift out by omniprobe needle and thinning. Lower currents ~30 pA were used for thinning from 200 nm to 100 nm and ~10 pA for < 50 nm to reduce the ion beam induced damage. Final cleaning was carried out at low kVs: 5 kV and 2 kV with minimum current (~7-14 pA) for removing the damaged amorphous layer formed during thinning for ~ 1 min each on both sides at 52 ± 3°, 5° and 7°.

**Acknowledgements:**

The device fabrication work was carried out at National Nanofabrication Centre (NNFC). The electrical and Raman characterization was carried out at Micro Nano Characterization Facility (MNCF) while the TEM facility was carried out at Advanced Facility for Microscopy and Microanalysis (AFMM). All these facilities are part of the Indian Institute of Science (IISc), Bengaluru. We acknowledge the help and support we received from the staff members of above-mentioned centres. We also acknowledge the meaningful insights we received from Arup Singha, from Department of Physics, IISc Bangalore. Authors would like to acknowledge Chandni Usha, from IAP, IISc Bangalore for allowing us to use her lab facilities. The authors acknowledge funding support received from Department of Science and Technology (DST COE) through piezoMEMS DST/TDT/AM/2022/084 and Science and Engineering Research Board (SERB-Core) through research grant (CRG/2022/003506).

**References**

1. Xue, F. *et al.* Room-Temperature Ferroelectricity in Hexagonally Layered α-$In_2Se_3$ Nanoflakes down to the Monolayer Limit. *Adv Funct Mater* **28**, (2018).

2. Bai, L. *et al.* Intrinsic Ferroelectric Switching in Two-Dimensional α- $In_2Se_3$. *ACS Nano* (2024) doi:10.1021/acsnano.4c06619.

3. Zhou, Y. *et al.* Out-of-Plane Piezoelectricity and Ferroelectricity in Layered α- $In_2Se_3$ Nanoflakes. *Nano Lett* **17**, 5508–5513 (2017).

4. Cui, C. *et al.* Intercorrelated In-Plane and Out-of-Plane Ferroelectricity in Ultrathin Two-Dimensional Layered Semiconductor $In_2Se_3$. *Nano Lett* **18**, 1253–1258 (2018).

5. Ding, W. *et al.* Prediction of intrinsic two-dimensional ferroelectrics in $In_2Se_3$ and other $III_2$-$VI_3$ van der Waals materials. *Nat Commun* **8**, (2017).

6. Zheng, C. *et al. Room Temperature In-Plane Ferroelectricity in van Der Waals* $In_2Se_3$. https://www.science.org (2018).

7. Hou, P., Lv, Y., Zhong, X. & Wang, J. α- $In_2Se_3$ Nanoflakes Modulated by Ferroelectric Polarization and Pt Nanodots for photodetection. *ACS Appl Nano Mater* **2**, 4443–4450 (2019).

8. Si, M. *et al.* A ferroelectric semiconductor field-effect transistor. *Nat Electron* **2**, 580–586 (2019).

9. Yang, F. *et al.* Emerging Opportunities for Ferroelectric Field-Effect Transistors: Integration of 2D Materials. *Advanced Functional Materials* vol. 34 Preprint at https://doi.org/10.1002/adfm.202310438 (2024).

10. Kim, J. Y., Choi, M. J. & Jang, H. W. Ferroelectric field effect transistors: Progress and perspective. *APL Mater* **9**, (2021).

11. Lee, S. *et al.* Low-temperature processed beta-phase $In_2Se_3$ ferroelectric semiconductor thin film transistors. *2d Mater* **9**, (2022).

12. Mohta, N., Rao, A., Remesh, N., Muralidharan, R. & Nath, D. N. An artificial synaptic transistor using an α-$In_2Se_3$ van der Waals ferroelectric channel for pattern recognition. *RSC Adv* **11**, 36901–36912 (2021).

13. Chu, T. C. *et al.* Resistive Switching in α-$In_2Se_3$ Lateral Field-Effect Transistors. *ACS Nano* **19**, 15100–15108 (2025).

14. Hao, S. *et al.* Activating Silent Synapses in Sulfurized Indium Selenide for Neuromorphic Computing. *ACS Appl Mater Interfaces* **13**, 60209–60215 (2021).

15. Xue, F. *et al.* Gate-Tunable and Multidirection-Switchable Memristive Phenomena in a Van Der Waals Ferroelectric. *Advanced Materials* **31**, (2019).

16. Li, Y. *et al.* Orthogonal Electric Control of the Out-of-Plane Field-Effect in Two-Dimensional Ferroelectric α-$In_2Se_3$.

17. Wan, S. *et al.* Room-temperature ferroelectricity and a switchable diode effect in two-dimensional α-$In_2Se_3$ thin layers. *Nanoscale* **10**, 14885–14892 (2018).

18. Duan, R. *et al.* A van der Waals ferroelectric switchable diode with ultra-high nonlinearity factor. *Appl Surf Sci* **693**, (2025).

19. Dai, M. *et al.* Intrinsic Dipole Coupling in 2D van der Waals Ferroelectrics for Gate-Controlled Switchable Rectifier. *Adv Electron Mater* **6**, (2020).

20. Chanchal, N., Jindal, K., Tomar, M. & Jha, P. K. A Self-Selector and Self-Rectifying Charge-Trap-Based Resistive Switching Device Using $In_2Se_3$ Thin Films. *ACS Appl Electron Mater* **6**, 3742–3753 (2024).

21. Wang, X., Feng, Z., Cai, J., Tong, H. & Miao, X. All-van der Waals stacking ferroelectric field-effect transistor based on $In_2Se_3$ for high-density memory. *Science China Information Sciences* **66**, (2023).

22. Park, S., Lee, D., Kang, J., Choi, H. & Park, J. H. Laterally gated ferroelectric field effect transistor (LG-FeFET) using α-$In_2Se_3$ for stacked in-memory computing array. *Nat Commun* **14**, (2023).

23. Dutta, D., Mukherjee, S., Uzhansky, M. & Koren, E. Cross-field optoelectronic modulation via inter-coupled ferroelectricity in 2D $In_2Se_3$. *NPJ 2D Mater Appl* **5**, (2021).

24. Mech, R. K., Solanke, S. V., Mohta, N., Rangarajan, M. & Nath, D. N. $In_2Se_3$ Visible/Near-IR Photodetector with Observation of Band-Edge in Spectral Response. *IEEE Photonics Technology Letters* **31**, 905–908 (2019).

25. Hu, Y. *et al.* Flexible Optical Synapses Based on $In_2Se_3$/$MoS_2$ Heterojunctions for Artificial Vision Systems in the Near-Infrared Range. *ACS Appl Mater Interfaces* **14**, 55839–55849 (2022).

26. Modi, G. *et al.* Electrically driven long-range solid-state amorphization in ferroic $In_2Se_3$. *Nature* **635**, 847–853 (2024).

27. Xu, C. *et al.* Two-dimensional ferroelasticity in van der Waals β'- $In_2Se_3$. *Nat Commun* **12**, (2021).

28. Jiang, Y. *et al.* 2D ferroelectric narrow-bandgap semiconductor Wurtzite' type α- $In_2Se_3$ and its silicon-compatible growth. *Nat Commun* **16**, 7364 (2025).

29. Pacchioni, G. One material, three phases. *Nat Rev Mater* **8**, 7 (2023).

30. Han, W. *et al.* Phase-controllable large-area two-dimensional $In_2Se_3$ and ferroelectric heterophase junction. *Nat Nanotechnol* **18**, 55–63 (2023).

31. Xu, K. et al; Optical Control of Ferroelectric Switching and Multifunctional Devices Based on van Der Waals Ferroelectric Semiconductors. *Nanoscale* **12**, 23488-23496 (2020)

32. Lei, Z. *et al.* Ultrafast Photocurrent Hysteresis in Photoferroelectric α- $In_2Se_3$ Diagnosed by Terahertz Emission Spectroscopy. *Sci. Adv* vol. 11 https://www.science.org (2025).

33. Lyu, F. *et al.* Temperature-Driven α-β Phase Transformation and Enhanced Electronic Property of 2H α- $In_2Se_3$. *ACS Appl Mater Interfaces* **14**, 23637–23644 (2022).

34. Guo, J. *et al.* Femtosecond Laser Manipulation of Multistage Phase Switching in Two-Dimensional $In_2Se_3$ Visualized via an In Situ Transmission Electron Microscope. *ACS Nano* **19**, 13264–13272 (2025).

35. Zheng, X. *et al.* Phase and Polarization Modulation in Two-Dimensional $In_2Se_3$ via in Situ Transmission Electron Microscopy. *Sci. Adv* vol. 8 https://www.science.org (2022).

36. Ignatans, R., Damjanovic, D. & Tileli, V. Individual Barkhausen Pulses of Ferroelastic Nanodomains. *Phys Rev Lett* **127**, (2021).

37. Huang, Y. T. *et al.* Dynamic observation of phase transformation behaviors in indium(iii) selenide nanowire based phase change memory. *ACS Nano* **8**, 9457–9462 (2014).

38. Luo, Z. *et al.* Reversible electrochromism in α- $In_2Se_3$ through ferroelectric switching induced phase transition. *Sci China Mater* **68**, 906–911 (2025).

39. Zhang, J. *et al.* Interlayer reconstruction phase transition in van der Waals materials. *Nat Mater* **24**, 369–376 (2025).

40. Wu, Y. *et al.* Stacking selected polarization switching and phase transition in vdW ferroelectric α- $In_2Se_3$ junction devices. *Nature Communications* **15**, (2024).

41. Kim, J. H. *et al.* Room Temperature Negative Differential Resistance with High Peak Current in MoS$_2$/WSe$_2$ Heterostructures. *Nano Lett* **24**, 2561–2566 (2024).

42. Tao, X. & Gu, Y. Crystalline-crystalline phase transformation in two-dimensional In$_2$Se$_3$ thin layers. *Nano Lett* **13**, 3501–3505 (2013).

43. Wan, S., Peng, Q., Wu, Z. & Zhou, Y. Nonvolatile Ferroelectric Memory with Lateral β/α/β In$_2$Se$_3$ Heterojunctions. *ACS Appl Mater Interfaces* **14**, 25693–25700 (2022).

# Supplementary Information

# Field-induced reversible phase transition and negative differential resistance in In$_2$Se$_3$ ferroelectric semiconducting FETs

Jishnu Ghosh[1,#], Shubham Parate[1,#], Arup Basak[1], Binoy Krishna De[1], Krishnendu Mukhopadhyay[1], Abhinav Agarwal[1], Gopesh Kumar Gupta[2], Digbijoy Nath[1]*& Pavan Nukala[1,2]*

[1] Center for Nanoscience and Engineering, Indian Institute of Science, Bengaluru, India

[2] Department of Materials Engineering, Indian Institute of Science, Bengaluru, India

* Corresponding authors' email

pnukala@iisc.ac.in, digbijoy@iisc.ac.in

[#] These authors contributed equally

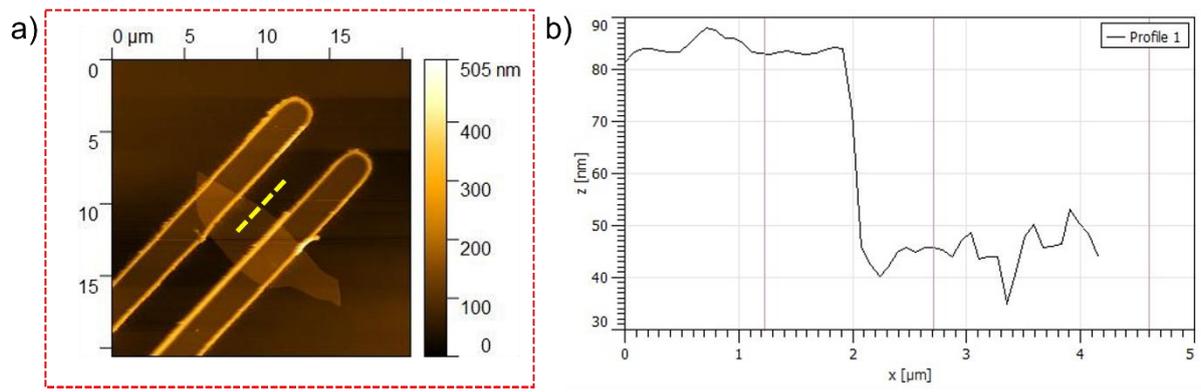

**Fig S1. AFM characterization:** a) AFM image of a few-layers In$_2$Se$_3$ flake. b) The step height measured along the indicated profile confirms the flake thickness to be 35 - 40 nm.

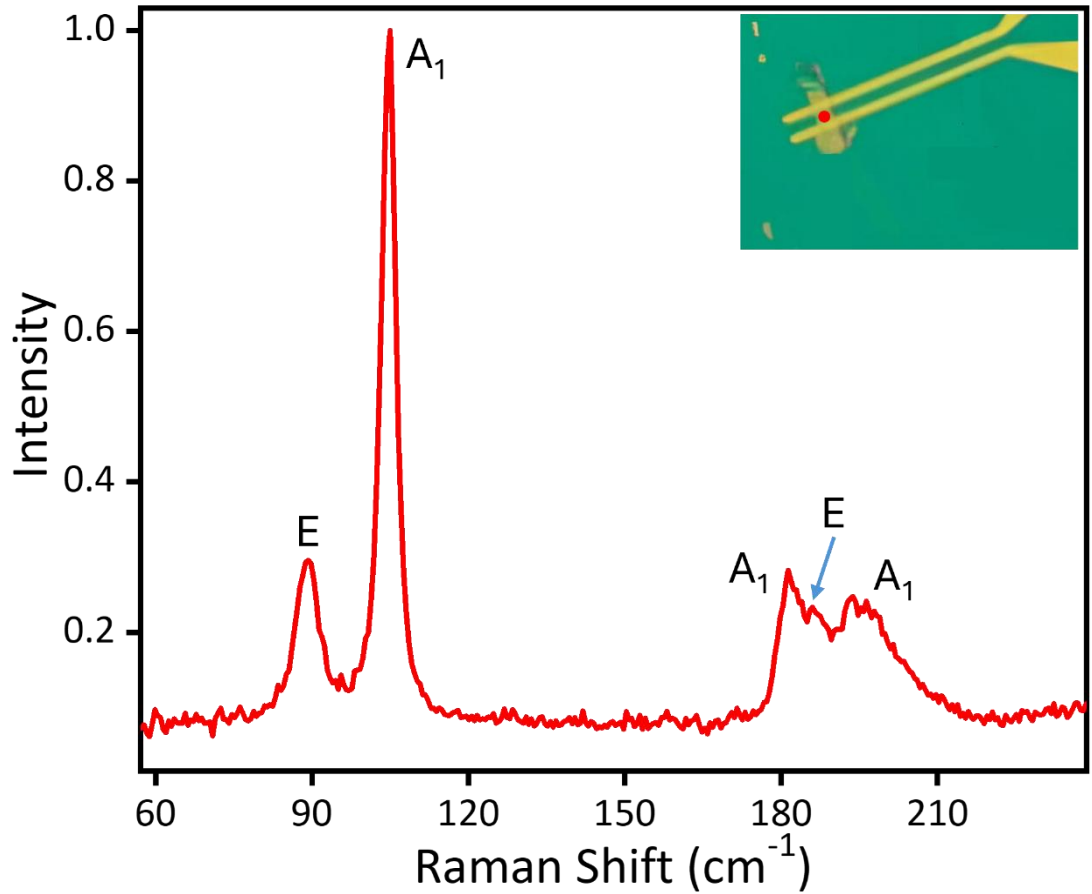

**Fig S2. Raman spectra.** Raman spectra of $In_2Se_3$ acquired from the channel of the FeS-FET. (Inset) shows the optical plan view image of the device, and the spectrum was acquired from the region marked in red dot.

**Supplementary note 1:** The Raman spectra of our $\alpha$−2H $In_2Se_3$ collected at the FeS-FET channel is shown in Fig S2. Peaks at 90, 185 cm$^{-1}$ are E Raman modes and peaks at 105, 181, 196 cm$^{-1}$ are $A_1$ Raman modes[1], and are consistent with $\alpha$−2H $In_2Se_3$

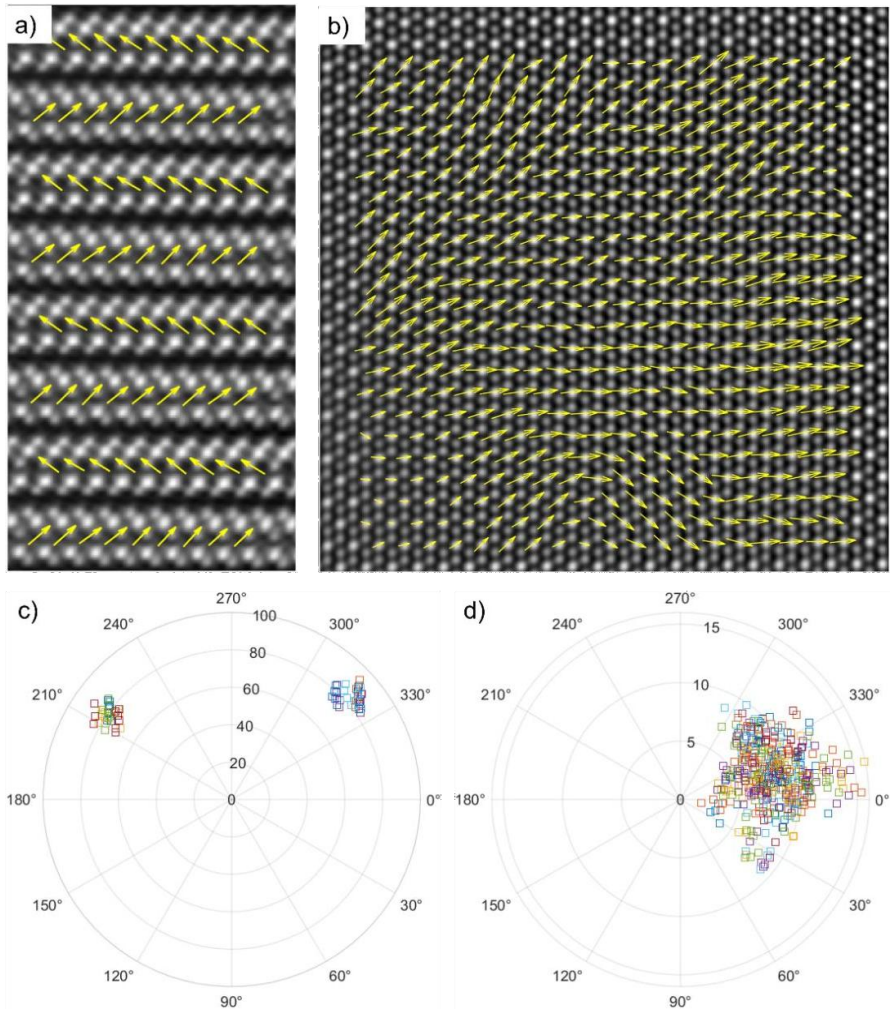

**Fig S3. Displacement mapping on pristine In$_2$Se$_3$:** (a) Cross sectional displacement mapping of Selenium with respect to the center of mass of unit cell of In$_2$Se$_3$. (b) Displacement mapping in plan-view showing lateral displacements of In from center of mass of Selenium hexagon. (c) Polar histogram capturing the displacement distribution in (a) and (d) showing distribution of displacements corresponding to (b). All the images correspond to the pristine state.

**Supplementary Note 2**: The lateral polarization from plan view and obtained from the cross-section polarization components are consistent. Average value in plan-view is ~7.8 pm with a standard deviation of ~3 pm while that obtained from cross section is ~6.6 pm which are close enough and very small compared to the average out of plane (c-axis) displacements of ~50 pm (Fig S3a).

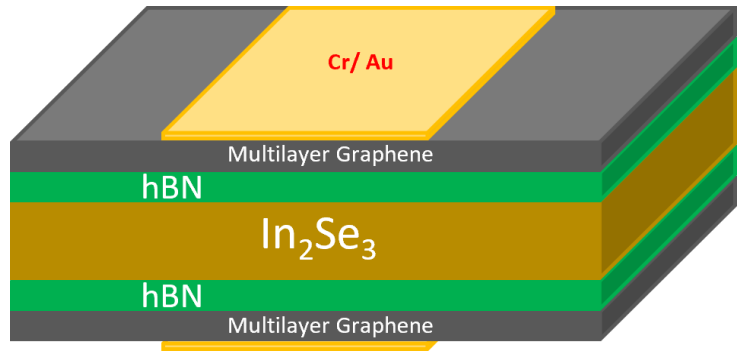

**Fig S4. Device Schematic:** Graphene/hBN/In$_2$Se$_3$/hBN/graphene stack for PE loop measurement

**Supplementary Note 3:** To check the ferroelectric switchable polarization, we performed P-E hysteresis loop measurement. The capacitor device structure is metal-insulator-semiconductor-insulator-metal (MISIM) hetero-structure, with In$_2$Se$_3$ embedded in hBN layers, and electrode with multilayer graphene (Fig S4). hBN reduces the leakage current, and passivates interfaces, which allows us to measure polarization switching in semiconducting In$_2$Se$_3$ (data presented in Fig 1f).

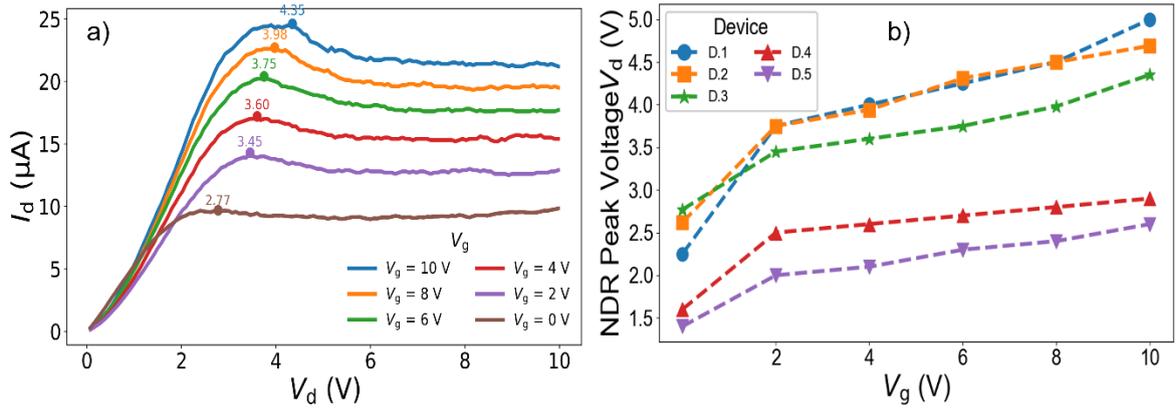

**Fig S5. NDR peak as a function of $V_g$:** (a) Drain voltages at the onset of the negative differential resistance (NDR) region. (b) Device to device NDR peak variation with respect to $V_g$

**Supplementary Note 4:** The value of drain voltage at onset of NDR regime shifts toward higher drain voltages ($V_d$) starting from 2.77 V to 4.35 V, as the gate voltage ($V_g$) increases from 0 V to 10 V. This trend indicates that the onset of NDR occurs at progressively larger $V_d$ values with increasing $V_g$. The orthogonal electric field generated by the gate bias enhances the stability of the polar α-2H phase, thereby suppressing in-plane transitions to a non-polar β phase. Consequently, NDR appears at higher drain voltages, indicating that a stronger in-plane field is required to induce the transition. Notably, while field induced stabilization of polar phases is common in ferroelectrics, in 2D ferroelectrics, it is only the out-of-plane field that stabilizes the polar phase, but in-plane field favors transitions to non-polar phase. In-plane field has the effect of disordering both structure and polarization, just like temperature does.

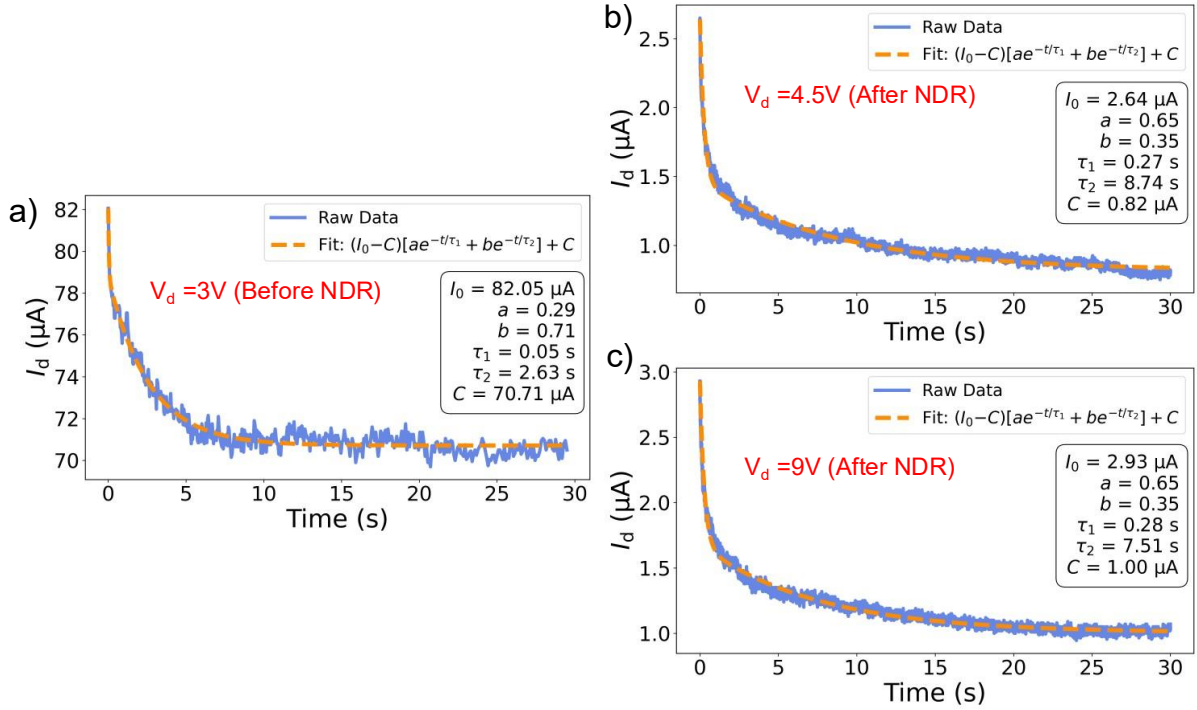

Fig S6. Temporal evolution of drain current before and after onset of NDR: DC measurements were conducted on the FeS-FET device to investigate the temporal evolution of drain current over a 30-second duration. In case 1 a), the drain current ($I_d$) was monitored at a fixed drain voltage of 3 V, capturing the behavior before the negative differential resistance (NDR) onset (LRS). In case 2 b) and c), measurements were taken at $V_d$ of 4.5 V and 9 V, respectively, both reflecting current dynamics following the NDR phenomenon (HRS). The resulting current versus time data were modeled using a standard bi-exponential fitting procedure to determine the relevant time constants.

**Supplementary Note 5:** The observed increase in the relaxation time constant following the onset of NDR suggests the involvement of slow, voltage-induced structural or electronic changes within the FeS-FET device. The transition from shorter (2 - 2.5 s) to longer (~8 s) relaxation time ($\tau_2$) points to a reversible phase transformation, likely from a low-resistance state (LRS) to a high-resistance state (HRS), driven by the applied drain voltage.

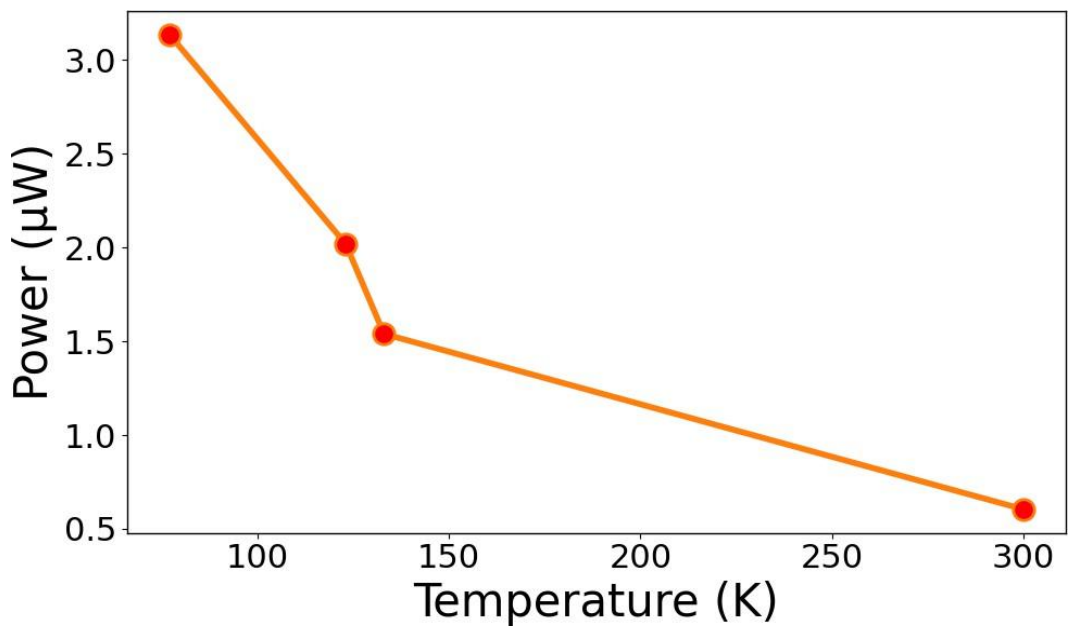

**Fig S7. Input power for the onset of NDR at different temperatures:** The amount of input power consumed by the FET on verge of NDR peaks recorded at different temperatures (77 K, 123 K, 133 K, 300 K)

**Supplementary Note 6:** NDR is consistently observed across all measured temperatures. The VI power input at the onset of NDR displays a modest decreasing trend with increasing ambient temperature. At higher temperatures, part of the energy required for the phase change is supplied by thermal energy, reducing the device's reliance on electrical input power.

**Supplementary note 7:**

**Electrothermal (COMSOL) simulations**

First, a geometry of the structure on which the electrothermal simulations is shown in Fig S8a. In addition to the dimensions mentioned in the figure, the In$_2$Se$_3$ flake and the Au pads have thickness of 25 nm each.

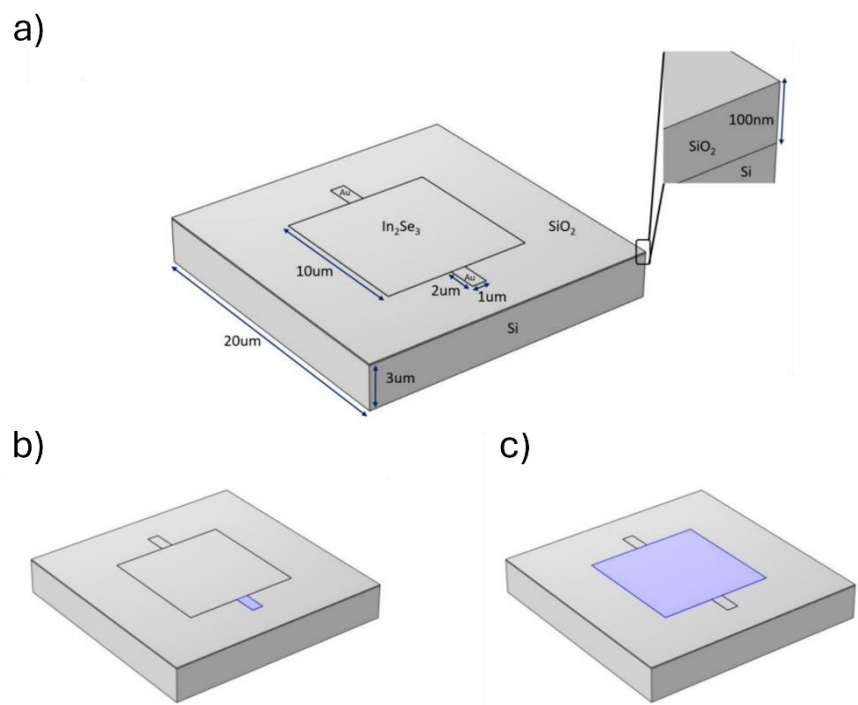

**Fig S8. Device geometry and probe setup**: a) Geometry of modelled device showing In$_2$Se$_3$ flake on SiO$_x$ substrate. Schematic showing b) voltage terminal and c) temperature probe.

In COMSOL, we track the value of any variable using a 'probe'. Temperature probe is used to keep track of the maximum temperature in the flake region. Schematic of the voltage terminals and temperature probe is shown in Fig S8b and Fig S8c respectively.

The materials properties used in the simulation are shown in Table S1. Electrical conductivity of In$_2$Se$_3$ flake was taken from experimental data which is shown in Table S2. The electrical conductivity of the flake (in LRS) was calculated from the experimental data. Temperature calculations were performed at the onset of NDR, where maximum power is delivered to the device. Experimental data at three different ambient temperatures (77 K, 135 K and room temperature) on several devices (see Table S2) were used to determine the VI power input to the system (at the onset of NDR).

**Table S1. Material properties:** Various material properties used in simulation.

| Physical Property | $In_2Se_3$ | $SiO_2$ | Si | Au |
|---|---|---|---|---|
| Electrical conductivity (S/m) | - | $10^{-19}$ | 10 | $45.6\times10^6$ |
| Thermal conductivity (J/m-K) | 15.6 | 1.5 | 130 | 317 |
| Density (Kg/m$^3$) | 5420 | 2650 | 2329 | 19300 |
| Specific heat $C_p$ (J/Kg-K) | 277 | 730 | 700 | 129 |
| Relative permittivity | 17 | 4.6 | 11.7 | 6.9 |

**Table S2. Experimental $I_d$ – $V_d$ data corresponding to peak NDR from which conductivity is calculated**

| T (K) | $I_d$ (µA) | $V_d$ (V) | Power (µW) | Conductivity (mS/m) |
|---|---|---|---|---|
| 77 | 3.13 | 1.15 | 3.61 | 0.147 |
| 77 | 2.04 | 1.3 | 2.66 | 0.253 |
| 123 | 1.83 | 1.55 | 2.85 | 0.337 |
| 123 | 0.85 | 1.4 | 1.19 | 0.659 |
| 133 | 4.71 | 1.725 | 8.13 | 0.146 |
| 133 | 1.43 | 1 | 1.43 | 0.278 |
| 133 | 1.21 | 1.35 | 1.64 | 0.443 |
| 300 | 0.38 | 1.8 | 0.68 | 1.895 |
| 300 | 0.34 | 1.7 | 0.57 | 2.019 |
| 300 | 0.33 | 1.7 | 0.55 | 2.068 |

With the constant voltage applied (determined by experimental onset of NDR), the simulation is run from t = 0 s to t = 10 s. Simulated steady state temperature profiles for representative devices at tested at ambient (background) temperatures ($T_o$) of 133 K and 300 K are shown in Fig S9 a, b.

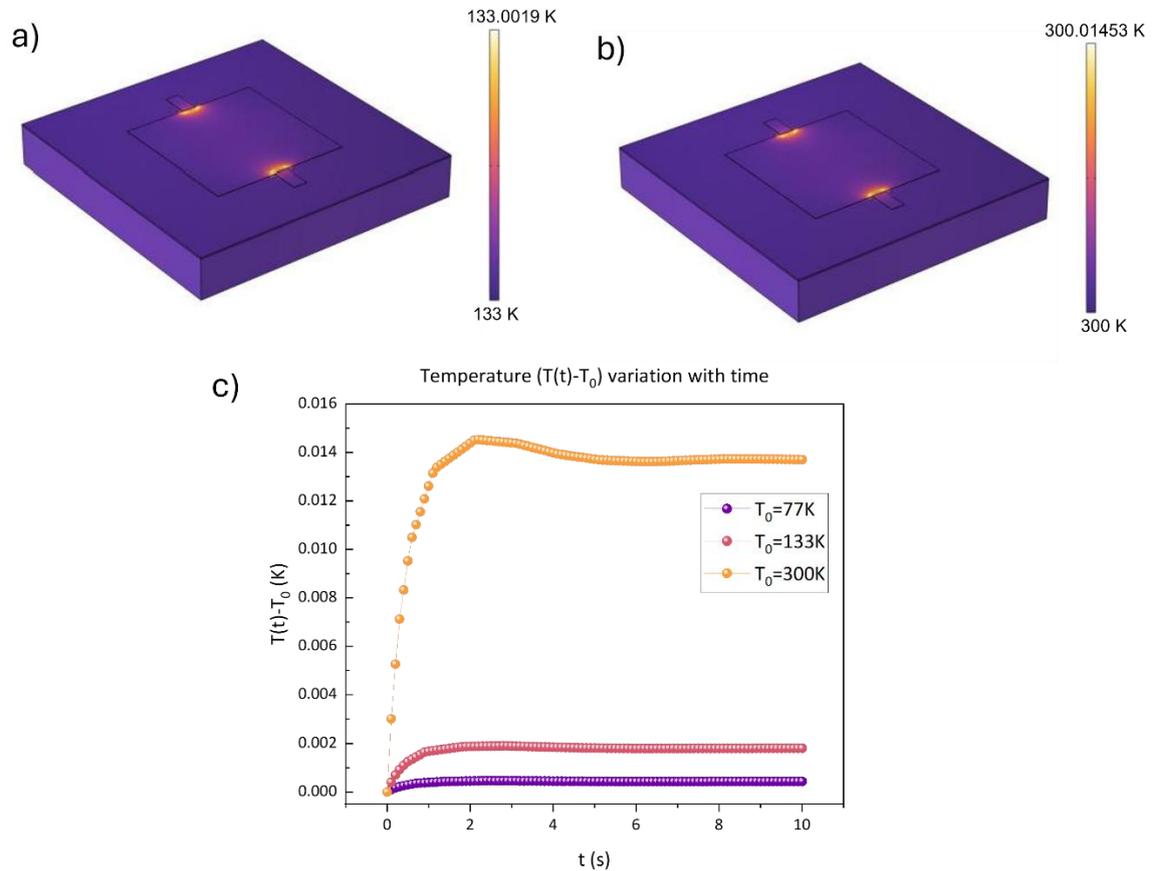

Fig S9. **Input voltage and output temperature:** a-b) Simulated temperature at a constant bias at 133 K and 300 K. c) Variation of temperature with time.

The evolution of temperature with time (T(t)) at three different $T_0$'s (77 K, 133 K, 300 K) evaluated near the contact where the temperature is maximum, is shown in Fig S9c. We see that temperature rise above the ambient in all the cases is almost nothing, ruling out device heating as a possible source of phase transition.

**Supplementary Note 8:** The average angle 2.66º for the Moiré periodicity was calculated by measuring the periodic occurrences of moiré pattern from the FFT (Fig 3d, FFT in inset in main text) and then feeding it into the relation[2]:

$$b = \frac{a}{2 \times Sin(\theta/2)} \qquad (1)$$

Where, $b$ = 8.6 nm is the average moiré periodicity evaluated from FFT, $a$ = 4.02 Å is lattice parameter, and $\theta$ is the twist angle between two layers.

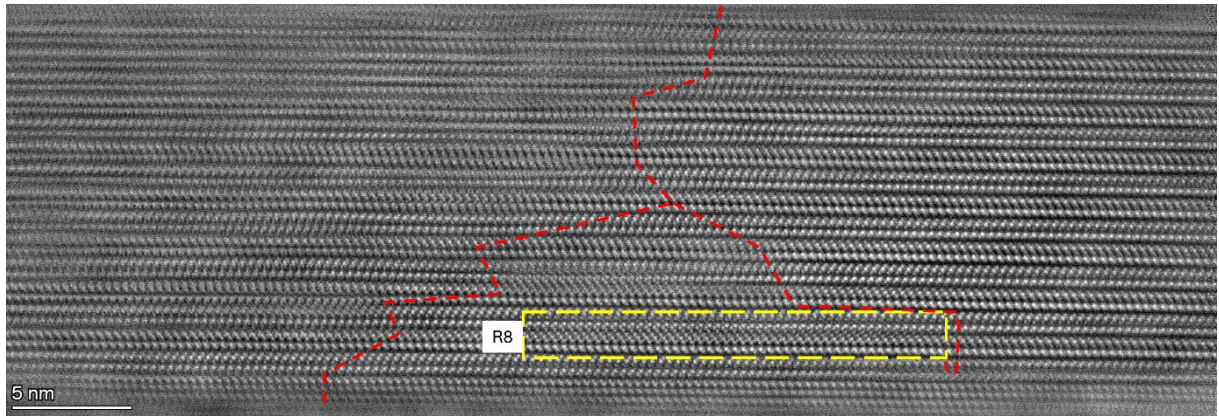

**Fig S10. Structural changes in a device cycled through NDR for a few times:** Large region HAADF image showing domains in focus (right) and out of focus (left) separated by red boundaries, and the middle domain represents a region undergoing α to β-like phase transition. The in-focus and out-of-focus domains possibly represent in-plane layer rotation corresponding to the Moiré patterns shown in the plan view. The yellow region is representative of a region clearly showing β-phase stacking (with Se-In-Se-In-Se being in a-b-c-a-b intralayer stacking).

**Supplementary Note 9**: We identify angles θ1- θ4 in a particular quintuple layer as quantifiable indicators for various phases in $In_2Se_3$. Table S3 shows the values of these indicators for quintuple layer orientations in several polymorphs of $In_2Se_3$. In the 2H phases of interest for us, a particular angle of interest is θ2, which in the α-2H phase is 90º and in β phase reduces to ~60º. We can identify phases using these criteria as shown in Fig S11.

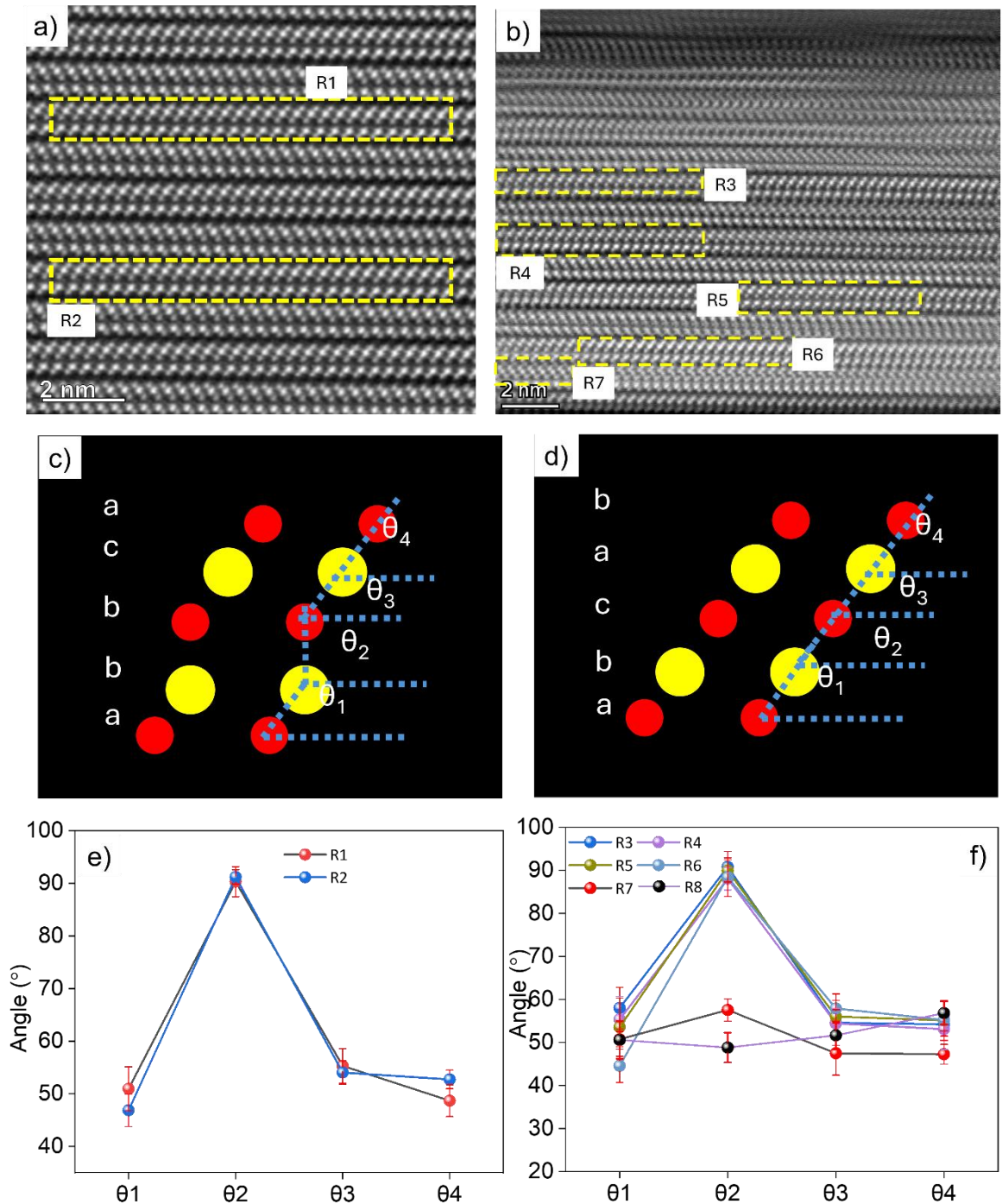

Fig S11. Quantifiable indicators of phase transitions: Cross-sectional images of pristine (a) and another cycled device (b) showing several marked regions from certain indicators have been extracted. These indicators are $\theta_1$- $\theta_4$ are shown in (c) for α-phase and (d) for β-like phase. These indicators are shown for various regions in the pristine α-phase (R1-R2) in (e) and in cycled device in (f) (R3-R7) and in Fig S10 (R8).

**Supplementary Note 10**: The pristine alpha phase quintuple orientation has a characteristic θ2 = 90º (Fig S11a). In Fig S11b (cycled device), however, while lot of regions still have values of the indicators corresponding to the α-phase (R3-R6), R7 and R8 corresponds to the indicators corresponding to β' phase (θ2 is close to ~60º).

**Table S3: θ1- θ4 indicators (as defined in S12c inset) for different phases.**

| Phase | Theta 1 | Theta 2 | Theta 3 | Theta 4 |
|---|---|---|---|---|
| α(3R) Exp. | 49.85 ± 2.62 | 89.76 ± 0.869 | 56.72 ± 1.52 | 50.5 ± 3.25 |
| β (2H-high temperature) Exp. | 59.34 ± 2.91 | 61.25 ± 2.46 | 55.08 ± 2.38 | 60.88 ± 3.17 |
| β'(1T) | 47 | 60 | 60 | 49 |
| α (2H) | 46 | 90 | 54 | 50 |

*Exp. refers to experimentally determined through HAADF-STEM images.

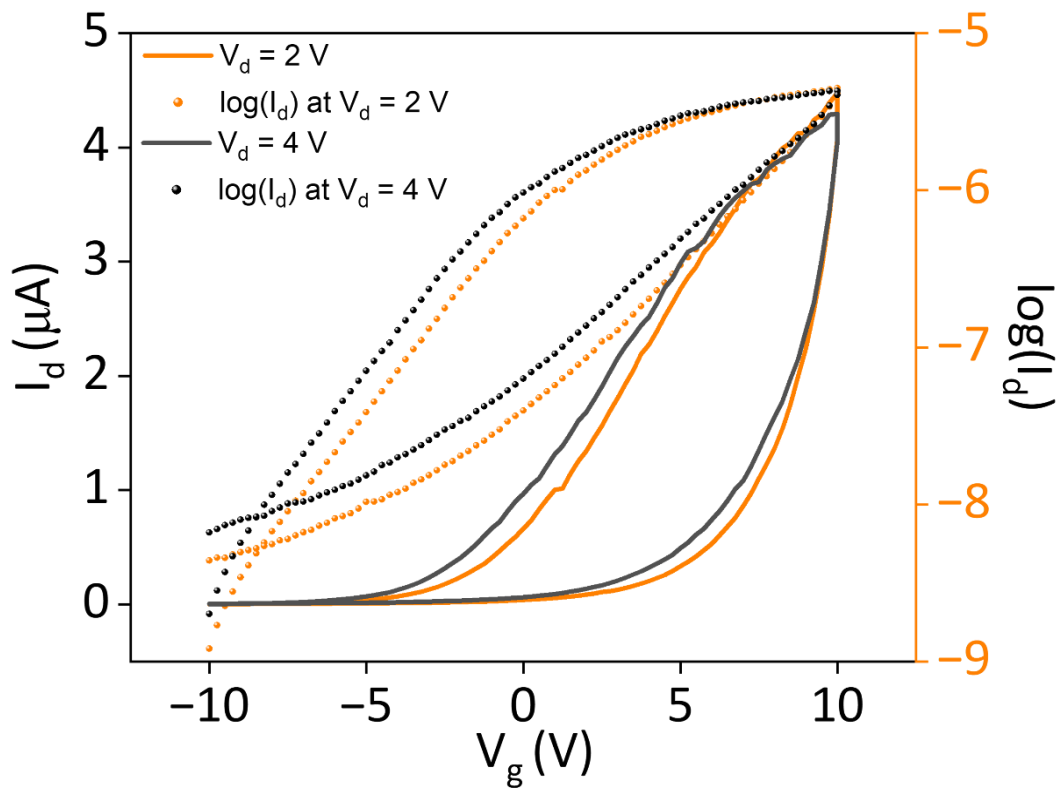

**Fig S12. Transfer characteristics of the FeS-FET:** $I_d$-$V_g$ characteristics showing clockwise hysteresis in LRS ($V_d$ = 2 V) and in HRS.

**Supplementary Note 11**: As-transferred $In_2Se_3$ flakes show an up polarization (Fig S3a). Transfer characteristics show a clockwise hysteresis, with the device naturally ON at $V_g$=0, possibly due to the polarization screening charges on the channel even at $V_g$=0. The transfer characteristics are also not very good, suggesting that our FeS-FETs are not good memory devices.

**Supplementary References**

1. Liu, L.; Dong, J *et. al.*; Atomically Resolving Polymorphs and Crystal Structures of In2Se3, *Chem. Mater.* **31**, 10143 (2019)

2. Liao, Y., Cao, W., Connell, J. *et al.* Evolution of Moiré Profiles from van der Waals Superstructures of Boron Nitride Nanosheets. *Sci Rep* **6**, 26084 (2016).